\documentstyle[aps,floats,twocolumn]{revtex}

\begin{document}

\draft

\twocolumn[\hsize\textwidth\columnwidth\hsize\csname@twocolumnfalse\endcsname 

\renewcommand{\thefootnote}{\fnsymbol{footnote}}

\title{Simulations of the 3d Ginzburg-Landau Model with Soft Amplitudes}

\author{Philippe Curty and Hans Beck} 

\address{Universit\'e de Neuch\^atel, 2000 Neuch\^atel, Switzerland}

\maketitle

\begin{abstract}
  { Using cluster Monte Carlo simulations, the 3d complex Ginzburg-Landau model reveals a first order transition when the amplitude $|\psi|$ of the complex field $\psi$ is sufficiently soft, i.e. adapts itself to the phase configurations of the field. This transition is driven by phase fluctuations in agreement with a previous analytical approach.}\\
  
\end{abstract}
]


The 3d complex Ginzburg-Landau (GL) model (for a review see \cite{goldenfeld,tinkham}) originally introduced by Landau to describe general features of phase transitions, is still of great relevance in domains like superconductivity, liquid crystals or particle physics, for example. In particular, the role of amplitude fluctuations and the order of the transition is still a subject of debate especially when one allows large amplitude fluctuations. The $\varepsilon$-expansion \cite{wilsonfisher} in the framework of the renormalisation group predicts that the 3d GL model should belong to the same universality class as the 3d XY model. On the other hand, a recent variational approach \cite{curty} has revealed the possibility of a first order transition induced by phase fluctuations. In this scenario, the interplay between phase and amplitude leads the amplitude to adapt itself to the configuration of phases and vice-versa. When the amplitude is soft, then the adaptation is so strong that a first order transition replaces the usual continuous transition.

   When two variables are coupled, it is well-known that fluctuations can change the order of the transition. For example, fluctuations of the magnetic field change the GL transition to a first order transition for type I superconductors \cite{halperin}. The three state Potts model in two dimensions is an opposite example: mean field theory predicts a first order transition, whereas the actual transition is continuous. For the GL model, it has been predicted \cite{curty} that the transition could be first order if the amplitude $|\psi|$ of the field $ \psi = |\psi| e^{i \phi}$ is sufficiently soft. If this is the case, phase fluctuations would drive the transition via the amplitude to a first order transition. It has been shown \cite{curty} that amplitudes are dominated by phase fluctuations in the domain of the first order transition for $d=3$. Therefore amplitude fluctuations can be neglected in this regime and phase fluctuations drive the XY transition to a first order transition via their action on the amplitude.

     Bormann and Beck \cite{bormann} have also shown that amplitude fluctuations might alter the cooperative phenomenon occurring with phases, in particular in dimension 2. Like the XY model, corresponding to a fixed value of the amplitude, the 2$d$GL model can be mapped onto a Coulomb gas describing vortex-antivortex pairs. As soon as one allows for amplitude variations, these topological excitations become energetically more favorable. Taking into account gaussian amplitude fluctuations, Bormann and Beck \cite{bormann} have shown that the system may be driven into a regime where - according to Minnhagen's phase diagram \cite{minnhagen} - a first order transition replaces the usual Kosterlitz-Thouless scenario.

    Monte Carlo simulations have become increasingly important to study statistical systems. Great progress has been made by using non-local algorithms for spin systems. Swendsen and Wang \cite{swendsen} have used the Fortuin-Kastelyn \cite{fortuin} percolation mapping for the Potts model to define domains or clusters to be inverted with zero free energy cost. Cluster Monte Carlo simulations have a small critical slowing down comparing to standard Metropolis algorithms and therefore allow to study the critical region for large systems.

    Concerning the 3d GL model, to our knowledge, there is no systematic study of the influence of amplitude fluctuations. Most of the existing simulations (for example \cite{nguyen}) are made in a domain where the XY behaviour dominate the transition.
Here we have combined a standard Monte Carlo algorithm  for the amplitude  $|\psi|$, and a Wolff algorithm \cite{wolff} for the phase $\phi$ in the same spirit as for the $\Phi^4$ model \cite{brower} where the real field $\Phi$ is the product of its amplitude and of a discrete variable $s=\pm 1$. The cluster Monte Carlo algorithm allows to reduce the simulation time in the region where the transition is XY-like or weakly first order. Indeed in this regime, the correlation length is very large, and therefore, using a cluster algorithm reduces strongly the critical slowing down \cite{swendsen,brower}.

{The aim of this letter} is to verify the analytical prediction \cite{curty} of the first order nature of the GL transition when amplitudes are soft by using cluster Monte Carlo  simulations.\\


According to Ginzburg-Landau theory, we define the effective hamiltonian functional
\begin{equation}
H[\psi]=\int{d^d \! r  \left[  \alpha  \left|\psi \right|^2+ {b\over 2} \left|\psi \right|^4   + { \gamma \over 2} \left|  \nabla\psi \right|^2 \right] } \label{hamiltonian}
\end{equation} 
where $\alpha$, $b$ and $\gamma$ are coefficients derived from a microscopic model. We take $\alpha < 0$ so that the critical region can be reached by varying the temperature in the Boltzmann factor. We put our system on a lattice, with lattice spacing $\varepsilon$. In order to establish the phase diagram,  we normalize the hamiltonian by setting $ \tilde{\psi} = {\psi / \sqrt{|\alpha|/b}}$,  ${ \vec u} = { \vec r / \xi}$, where $ \xi^2= \gamma / |\alpha| $ is the mean field correlation length. The normalized hamiltonian is then:
\begin{equation}
  H[\tilde{\psi}] = k_B  \tilde{V}_0 \sum_{i=1}^{N}   \left[ {\tilde{\sigma} \over 2}  \left( | \tilde{\psi}_i |^2 - 1  \right)^2 + {1 \over 2} \sum_{\mu=1}^{d} \left|  \tilde{\psi}_i -\tilde{\psi}_{i+\mu} \right|^2 \right]
\end{equation}
where we have removed the constant term, and $\mu$ points to the $d$ nearest neighbours of the site $i$. $N$ is the total number of sites. Only two competing parameters remain:
$$
\tilde{\sigma} := \varepsilon^2 / \xi^2    \hspace{2 cm}           \tilde{V}_0 := {1 \over k_B} {|\alpha | \over b} \gamma  \varepsilon^{d-2} 
$$
$\tilde{\sigma}$ controls amplitude fluctuations. When $\tilde{\sigma}$ is large, then the amplitude $|\tilde{\psi}|$ is forced to take the value 1 to avoid too large costs for the energy and the only physics is then the one of the XY model. $\tilde{V}_0$ corresponds to the zero temperature phase stiffness and is proportional to the superfluid density $|\alpha|/b$ in the language of superconductivity. When $ \tilde{V}_0/T_c$ is large, the critical region is small and the system has the properties of the mean field theory. When $\tilde{V}_0/T_c$ is of the order of 1, phase fluctuations become very large and yield a critical temperature $T_{\phi}$ which is an upper bound for the true critical temperature $T_c$ \cite{emery}.

The canonical partition function $Z$ is then the trace over all possible configurations of the complex field $\tilde{\psi}$:
\begin{equation}
Z = \int D\tilde{\psi} \ e^{- H/\tilde{T}}
\label{partitionfunction}
\end{equation} 
where the reduced temperature is defined as $\tilde{T} = T/\tilde{V}_0$.\\


In the Swendsen-Wang algorithm, many clusters are generated and then flipped. Here, our idea is to use a standard Monte Carlo procedure for the amplitude $|\tilde{\psi}|$ and a Wolff \cite{wolff} algorithm for the phase $\phi$ as it has been done for the real $\Phi^4$ model \cite{brower}. A Wolff algorithm is similar to the Swendsen-Wang algorithm but only one single cluster is flipped at each step.  The detailed procedure is the following:
\begin{enumerate}

\item  With fixed amplitude $| \tilde{\psi}|$, introduce phase variables $\phi$ of the complex field $ \tilde{\psi}$, and take a  site $k$ (the seed) at random. Choose a random direction $\bf z$ with $|{\bf z}|=1$. For a site $i$, we define a unit vector spin ${\bf s}_i = \{ $Re$ [ \tilde{\psi}_i ], $Im$ [  \tilde{\psi}_i]  \} / | \tilde{\psi}_i|$. Nearest neighbours ${\bf s}_j$ of the cluster can be added to the cluster only if the component ${\bf s}_j \cdot {\bf z}$ has the same sign as the seed component ${\bf s}_k \cdot {\bf z}$ (i.e. they point to the same half plane perpendicular to $\bf z$).  Then, if ${\bf s}_i$ is already in the cluster and a nearest neighbour ${\bf s}_j$ of ${\bf s}_i$ points the same half plane, then ${\bf s}_j$ is added to the cluster with probability
\begin{equation}
p = 1 - \exp \left[ 2  ( {\bf z} \cdot {\bf s}_i ) ({\bf z}  \cdot   {\bf s}_j ) \right]
\label{probability}
\end{equation}
If the cluster stops growing, the process is interrupted and  all spins of the cluster are flipped with respect to the axis perpendicular to $\bf z$:
\begin{equation}
{\bf s}_i \ \rightarrow \ {\bf s}_i - 2  ( {\bf z} \cdot {\bf s}_i )  {\bf z}
\label{flip}
\end{equation}
Considering two configurations $\mu$ and $\nu$ with an energy difference $\Delta H$, the ratio of the selection probabilities $W(\mu \rightarrow \nu)$ and  $W(\nu \rightarrow \mu)$ is then equal to $e^{-\beta \Delta H}$, and the detailed balance is fulfilled (see the article of U. Wolff for the proof \cite{wolff}).

\item  Adapt the amplitude $|\tilde{\psi}|$ to the phase configuration with a standard Metropolis Monte Carlo algorithm.

\end{enumerate}



\begin{figure}[h]

\let\picnaturalsize=N
\def\picsize{8.5 cm}
\def\picfilename{Fig1_Comp.eps}
\ifx\nopictures Y\else{\ifx\epsfloaded Y\else\input epsf \fi
\let\epsfloaded=Y
\centerline{\ifx\picnaturalsize N\epsfxsize \picsize\fi \epsfbox{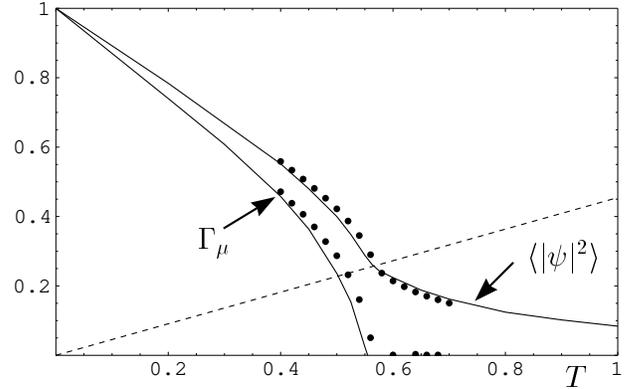}}}\fi

\caption{Comparison between standard MC simulations (lines) and cluster MC simulations (points).($\alpha =18 (T - 1)$, $b = 18$)}
\label{ComparisonSudbo}
\end{figure}


We present now the results of the cluster Monte Carlo simulations that have been performed on the isotropic 3d GL model with a complex field $\psi$. The expectation value of a  variable $A$ is defined by:
\begin{equation}
\left< A \right> = {1 \over M} \sum_{m=1}^M  A_m
\label{expectationvalue}
\end{equation}
where $M$ is the number of MC sweeps, and $A_m$ the value of $A$ at the end of the sweep $m$. One sweep consists of points 1) and 2) explained above.

The first measured quantity is the mean square amplitude:
\begin{equation}
\left< |\tilde{\psi}|^2 \right> = {1 \over N} \left< \sum_i  | \tilde{\psi}_i|^2  \right>
\label{meanpsi2}
\end{equation}
In order to determine the critical temperature, the helicity modulus is also computed:
\begin{eqnarray}
 \lefteqn{ \Gamma_{\mu}  =  {1 \over N} \left< \sum_i  |\tilde{\psi}_i|  |\tilde{\psi}_{i+\mu}| \cos ( \phi_i - \phi_{i+\mu}) \right>   } \nonumber \\
& & \qquad \quad  - {1 \over N \tilde{T}} \left<  \left[ \sum_i  |\tilde{\psi}_i|  |\tilde{\psi}_{i+\mu}| \sin ( \phi_i - \phi_{i+\mu}) \right]^2 \right>
\label{helicitymodulus}
\end{eqnarray}
The helicity modulus $\Gamma_{\mu}$ is a measure of the correlation of phases in the direction $\mu$. $\Gamma_{\mu}$ is zero above $T_c$ and different from zero below $T_c$.

A quantity of interest is also the XY energy of the phases. As in the XY model it is:
\begin{equation}
\left< f \right>  = {1 \over N} \left< \sum_i  f_i \right>
\label{meanpsi2}
\end{equation}
where $f_i= \sum_{\mu} [ 1 - \cos (\phi_i-\phi_{i+\mu})]/3$ is the normalised XY energy.

Using different lattice sizes, enough sweeps have been used in order to have error bars of the order of the symbols used to plot the curves. In practice, the thermal equilibration was reached between 200 and 1000 sweeps, and then $10^4$ sweeps were sufficient to have the required precision for $N = 10^3$. We used lattice sizes from $3^3$ to $15^3$ which is enough to reveal the first order transition. For point 2), we let $|\tilde{\psi}|$ fluctuate  from $0$ to a limit $L$ that is chosen so that the probability of the least probable configuration is less than $10^{-10}$. This choice gives an acceptance rate of new amplitude configurations that is approximately between 30 \% and 50 \%.

In order to control the validity of this method of cluster Monte Carlo simulations, we compare it with standard Monte Carlo simulations performed on the same model by Nguyen and Sudb\o \cite{nguyen}. In figure \ref{ComparisonSudbo}, one can see that although the system sizes are different, $15^3$ for the cluster MC method and $60^3$ for the other one, the two methods give results that are quite close.


\begin{figure}

\let\picnaturalsize=N
\def\picsize{8.5 cm}
\def\picfilename{Fig2_Ampl.eps}
\ifx\nopictures Y\else{\ifx\epsfloaded Y\else\input epsf \fi
\let\epsfloaded=Y
\centerline{\ifx\picnaturalsize N\epsfxsize \picsize\fi \epsfbox{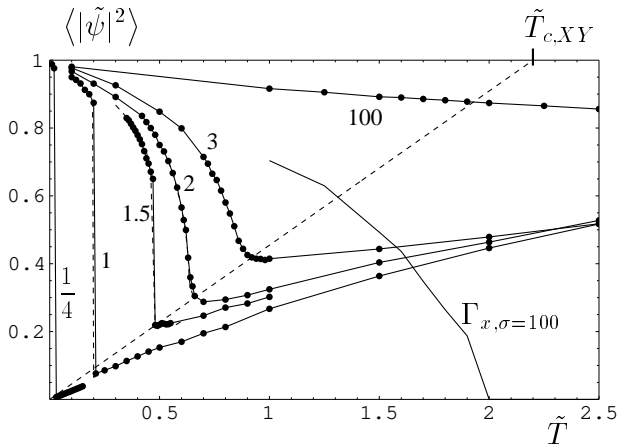}}}\fi

\caption{{  Reduced mean amplitude for $ d = 3$ for different values of $\tilde \sigma$.} { Points are for $N=10^3$ and dashed lines for $N=15^3$.  The slanting dashed line shows the XY transition ending at $\tilde{T}_{c,XY} \approx 2.2 $ ($\tilde \sigma \rightarrow \infty$).}}
\label{amplitudeSigma}
\end{figure}


Results for different $\tilde{\sigma}$ are shown in figure \ref{amplitudeSigma} for $N=10^3$ (points) and $N=15^3$ (dashed lines). The mean square amplitude $\langle |\tilde{\psi}|^2 \rangle $ is shown as a function of the reduced temperature $\tilde{T}$. The intersection between $\langle |\tilde{\psi}|^2 \rangle $ and the slanting dashed line shows the XY critical temperature as it would be obtained with the XY model and a non-fluctuating coupling constant $K(\tilde{T})= \langle |\tilde{\psi}|^2 \rangle /\tilde{T}$. At $T_c$, the we have the equation:
 \begin{equation}
 K(\tilde{T}_c)= \langle |\tilde{\psi}|^2 \rangle (\tilde{T}_c)/\tilde{T}_c = \tilde{T}_{c,XY}.
\label{equationTc}
\end{equation}
In order to control the limiting case of the XY model, simulations have been made for $\tilde{\sigma}=100$: the critical temperature $\tilde{T}_c$ (where $\Gamma_{\mu}=0$) is near the XY critical temperature $\tilde{T}_{c,XY} \approx 2.2$, and moreover $\tilde{T}_c$ obeys equation (\ref{equationTc}). The most striking feature is the appearance of a first order transition below a critical value $\tilde{\sigma}^* \approx 2$. If $\tilde{\sigma} > \tilde{\sigma}^* $, there is no observable first order transition and the transition is completely driven by the phase as also shown by Nguyen and Sudb\o \cite{nguyen}. When $\tilde{\sigma} < \tilde{\sigma}^* $, then $ < |\tilde{\psi}|^2 >$ exhibits a jump at the critical temperature $\tilde{T}_c$. Simulations made for larger systems also confirm  the presence of this critical point  $ \tilde{\sigma}^*$. Indeed, results for $N=15^3$ are very close to the curves obtained for  $N=10^3$.
We have also used standard Metropolis Monte Carlo simulations for the regime where the transition is strongly first-order ($ \tilde{\sigma}<< \tilde{\sigma}^*$). The same results are obtained when comparing to the cluster Monte Carlo method although the Metropolis algorithm takes much longer simulation time than the cluster algorithm even in the first order regime.

The thermal equilibration is faster for systems with small $\tilde{\sigma}$. Therefore, we can exclude the possibility that the Monte Carlo trajectory in the phase space jumps to another value if one would take much more MC sweeps.


\begin{figure}[h]

\let\picnaturalsize=N
\def\picsize{8.5 cm}
\def\picfilename{Fig3_Distrib.eps}
\ifx\nopictures Y\else{\ifx\epsfloaded Y\else\input epsf \fi
\let\epsfloaded=Y
\centerline{\ifx\picnaturalsize N\epsfxsize \picsize\fi \epsfbox{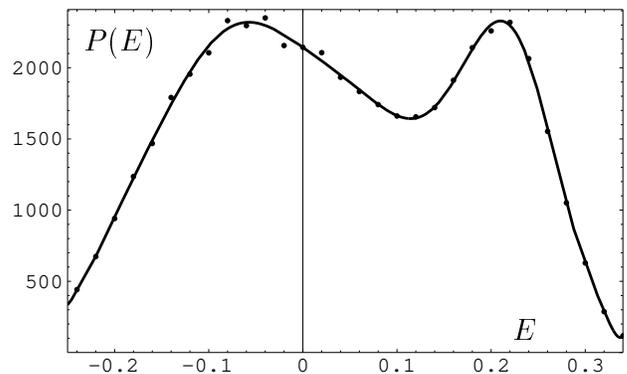}}}\fi

\caption{  Energy probability distribution near $\tilde{T}_c$ as a function of the energy $E$. ($\tilde{\sigma}=1.5$,  $\tilde{T}=0.4945$, $N=4^3$) }
\label{EnergyDistribution}
\end{figure}


A typical plot of the energy distribution $P(E)$ near $\tilde{T}_c$ is shown in figure \ref{EnergyDistribution} when the transition is first-order ($\tilde{\sigma} = 1.5$). The two peaks of the distribution have the same height, and  this reveals two coexisting phases near $\tilde{T}_c$. This plot was obtained by sampling 50000 sweeps for $N=4^3$. Larger system sizes only emphasize the two peaks. However, it is difficult to control the shape of the peaks because the first order transition is not due a fundamental degeneracy in the hamiltonian like for the Potts model \cite{lee}. These two peaks disappear when the transision is continuous unlike the Potts model so that it is not possible to determine the critical point $\tilde{\sigma}^*$ by using the sampling method of Lee and Kosterlitz \cite{lee}.

\begin{figure}

\let\picnaturalsize=N
\def\picsize{8.5 cm}
\def\picfilename{Fig4_Anticor.eps}
\ifx\nopictures Y\else{\ifx\epsfloaded Y\else\input epsf \fi
\let\epsfloaded=Y
\centerline{\ifx\picnaturalsize N\epsfxsize \picsize\fi \epsfbox{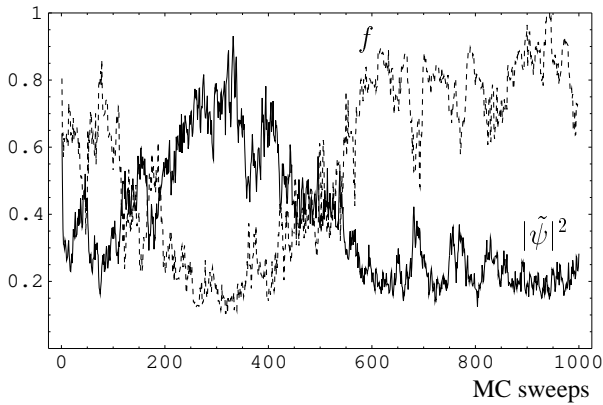}}}\fi

\caption{Monte Carlo trajectories of $ |\tilde{\psi}|^2 $ (thick line) and $f_i$ (dashed line) showing the anti-correlation between these two variables (the parameters are the same as in figure \protect{\ref{EnergyDistribution}}). }
\label{AntiCorrelation}
\end{figure}

An interesting feature is the anti-correlation between $\langle |\tilde{\psi}|^2 \rangle$ and  $\langle f \rangle$ during the MC simulation (see figure \ref{AntiCorrelation}). This anti-correlation is stronger at $\tilde{T}_c$ and is due to the coupling between amplitudes and phases.

\begin{figure}

\let\picnaturalsize=N
\def\picsize{8.5 cm}
\def\picfilename{Fig5_Jump.eps}
\ifx\nopictures Y\else{\ifx\epsfloaded Y\else\input epsf \fi
\let\epsfloaded=Y
\centerline{\ifx\picnaturalsize N\epsfxsize \picsize\fi \epsfbox{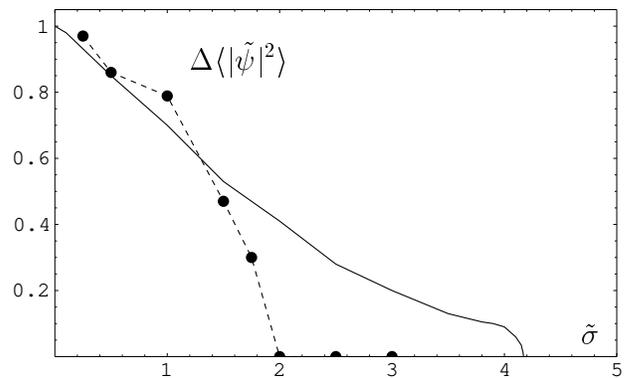}}}\fi

\caption{{ Size of the amplitude discontinuity as a function of $\tilde \sigma$.} We compare the simulations (points) with the analytical result of reference \protect{\cite{curty}}. }
\label{GraphJump}
\end{figure}

In figure \ref{GraphJump}, the value of the mean amplitude jump $ \Delta \langle |\tilde{\psi}|^2 \rangle$  is plotted as a function of the parameter $\tilde{\sigma}$. The analytical result from reference \cite{curty} is also shown (continuous line). Our calculations show that by increasing the size of the system the simulations get closer to the analytical result. However, a precise value of $\tilde{\sigma}^*$ is difficult to determine because of the finite size effects that round the jump at the transition. Nevertheless, we can say that $\tilde{\sigma}^*$ is restricted to the domain:
\begin{equation}
2 < \ \tilde{\sigma}^*  <  4.5
\label{sigmaDomain}
\end{equation}


In conclusion, cluster Monte Carlo simulations performed on the 3d complex Ginzburg-Landau model exhibit a first order transition for soft amplitudes when the parameter $\tilde{\sigma}$ is  approximately smaller than 2. These conclusions are in good qualitative agreement with a previous analytical work \cite{curty} that predicts a first order transition for $\tilde{\sigma} < 4.5$. Therefore, if we take into account the rounding of the jump at the transition in the simulations, the critical point $\tilde{\sigma}^*$, that separate first order and  continuous domains, should lie between $2$ and $4.5$.
In order to have a better accuracy and a precise value of the critical $\tilde{\sigma}^*$, one should use much larger systems and  finite size scaling.

Concerning a possible observation of this mechanism, the first order transition between nematic and smectic A 
seems to be a good candidate to measure the influence of phase fluctuations on the smectic A density $|\psi|$. It is believed that the first order nature of this transition is due to the HLM mechanism \cite{halperin}, but recent experiments \cite{yethiraj} show some differences between HLM predictions and measurements. It is therefore not clear which mechanism is responsible for the first oder nature of the transition. Same conclusions apply to the case of type I superconductors where however a latent heat has never been measured.

As already mentioned in \cite{curty}, high temperature superconductors near the overdoped regime appear to be good candidates for an observation of a first order transition, as for example the cuprate Bi$_2$Sr$_2$CaCu$_2$O$_8$.\\

 We thank H. Fort for useful discussions. This work has been supported by the Swiss National Science Foundation (project No. 2000-056803.99).


\end{document}